\documentclass[floatfix,twocolumn,showpacs,preprintnumbers,amsmath,amssymb,prr,superscriptaddress,longbibliography, nofootinbib]{revtex4-1}
\usepackage{color}
\usepackage[usenames,dvipsnames,svgnames,table]{xcolor}
\usepackage[colorlinks=true,linkcolor=blue,urlcolor=blue,citecolor=blue]{hyperref}
\usepackage{mathtools}
\usepackage{graphicx}
\usepackage{dcolumn}
\usepackage{array}
\usepackage{lipsum}
\usepackage{bm}
\usepackage{subfigure}
\usepackage{amssymb}
\usepackage{multirow}
\usepackage{tabularx}
\usepackage{amsmath}
\usepackage{braket}
\usepackage{csquotes}
\graphicspath{{plots/}}
 \usepackage{lipsum}
\usepackage{mathrsfs}
\usepackage{MnSymbol}
\usepackage{caption}
\usepackage{subcaption}
\newcommand{\beq}{\begin{equation}}
\newcommand{\eeq}{\end{equation}}
\newcommand{\bea}{\begin{eqnarray}}
\newcommand{\eea}{\end{eqnarray}}

\renewcommand{\vec}[1]{\mathbf{#1}}

\begin{document}
%\underline{Draft version:}
\title{ %Exact normalization
Applying the Liouville-Lanczos Method of Time-Dependent
Density-Functional Theory to Warm Dense Matter
}

\author{Zhandos~A.~Moldabekov}
\email{z.moldabekov@hzdr.de}

\affiliation{Center for Advanced Systems Understanding (CASUS), D-02826 G\"orlitz, Germany}
\affiliation{Helmholtz-Zentrum Dresden-Rossendorf (HZDR), D-01328 Dresden, Germany}

\author{Sebastian Schwalbe}

\affiliation{Center for Advanced Systems Understanding (CASUS), D-02826 G\"orlitz, Germany}
\affiliation{Helmholtz-Zentrum Dresden-Rossendorf (HZDR), D-01328 Dresden, Germany}

\author{Thomas Gawne}

\affiliation{Center for Advanced Systems Understanding (CASUS), D-02826 G\"orlitz, Germany}
\affiliation{Helmholtz-Zentrum Dresden-Rossendorf (HZDR), D-01328 Dresden, Germany}

\author{Thomas~R.~Preston}
\affiliation{European XFEL, D-22869 Schenefeld, Germany}

\author{Jan Vorberger}
\affiliation{Helmholtz-Zentrum Dresden-Rossendorf (HZDR), D-01328 Dresden, Germany}

\author{Tobias Dornheim}
% \email{t.dornheim@hzdr.de}

\affiliation{Center for Advanced Systems Understanding (CASUS), D-02826 G\"orlitz, Germany}
\affiliation{Helmholtz-Zentrum Dresden-Rossendorf (HZDR), D-01328 Dresden, Germany}

\begin{abstract}

\textit{Ab initio} modeling of dynamic structure factors (DSF) and related density response properties in the warm dense matter (WDM) regime is a challenging computational task. The DSF, convolved with a probing X-ray  beam and instrument function, is measured in X-ray Thomson scattering (XRTS) experiments, which allows for the study of electronic structure properties at the microscopic level. Among the various \textit{ab initio} methods, linear response time-dependent density functional theory (LR-TDDFT) is a key framework for simulating the DSF. The standard approach in LR-TDDFT for computing the DSF relies on the orbital representation. A significant drawback of this method is the unfavorable scaling of the number of required empty bands as the wavenumber increases, making LR-TDDFT impractical for modeling XRTS measurements over large energy scales, such as in backward scattering geometry. 
In this work, we consider and test an alternative approach to LR-TDDFT that employs the Liouville-Lanczos (LL) method for simulating the DSF of WDM. This approach does not require empty states and allows the DSF at large momentum transfer values and over a broad frequency range to be accessed. We compare the results obtained from the LL method with those from the solution of Dyson’s equation using the standard LR-TDDFT within the projector augmented-wave formalism for isochorically heated aluminum and warm dense hydrogen. 
Additionally, we utilize exact path integral Monte Carlo (PIMC) results for the imaginary-time density-density correlation function (ITCF) of warm dense hydrogen to rigorously benchmark the LL approach. We discuss the application of the LL method for calculating DSFs and ITCFs at different wavenumbers, the effects of pseudopotentials, and the role of Lorentzian smearing. The successful validation of the LL method under WDM conditions makes it a valuable addition to the \textit{ab initio} simulation landscape, supporting experimental efforts and advancing WDM theory.

\end{abstract}

\maketitle

\section{Introduction}

The physical and chemical processes occurring in matter at elevated temperatures and densities are of great interest due to their relevance in fields such as astrophysics~\cite{fortov_review}, material science~\cite{Kraus2016}, and inertial confinement fusion \cite{Batani_mre_2021,Betti2023,drake2018high}. The state of materials characterized by strongly correlated electrons at temperatures around the electron Fermi temperature is commonly referred to as warm dense matter (WDM) \cite{wdm_book, Bonitz_pop_2024, moldabekov2024review}.

Experiments designed to explore matter under extreme conditions are routinely conducted at various research centers, including the European XFEL in Germany \cite{Zastrau_2021,Rajan2023}, SLAC in the USA \cite{Fletcher2015}, SACLA in Japan \cite{Hara2013two}, the OMEGA laser facility in the USA \cite{Lees_prl_2021}, the Z Pulsed Power Facility (Z-machine) in Albuquerque, USA \cite{Knudson_2015}, and the National Ignition Facility in Livermore, USA \cite{Tilo_Nature_2023}.
The diagnostics of material properties in these experiments is a highly challenging task that often requires significant post-processing efforts to extract relevant information, and typically involves the use of computer simulations \cite{Dornheim_review}. An important example of such a diagnostic tool is X-ray Thomson scattering (XRTS) \cite{Glenzer_PRL_2007, Gregori2008, bellenbaum2024modelfreetemperaturediagnosticswarm}. When the shape of the probing X-ray beam are known (specifically, the combined source-and-instrument function \cite{Gawne_JAP_2024}), XRTS measurements can provide access to the dynamical structure factor (DSF) of the electrons, $S(q,\omega)$. This information enables one to infer key information such as the temperature \cite{Dornheim_T_2022,DOPPNER2009182}, Rayleigh weights \cite{dornheim2024_Rayleigh}, and mass density \cite{Tilo_Nature_2023,dornheim2024unravelingelectroniccorrelationswarm,dornheim2024_Rayleigh}. 

Interpreting the XRTS spectrum necessitates a theoretical analysis, which can be performed using straightforward and computationally efficient approximate models \cite{Gregori_PRE_2003}. However, these models often result in significant uncertainties in accuracy, both due to the dependence of conditions on the model used~\cite{boehme2023evidence} and the ability for a wide range of conditions to produce similarly plausible fits to the spectrum~\cite{Kasim_pop_2019}. Alternatively, high-fidelity (but computationally intensive) \textit{ab initio} calculations can be employed, such as Kohn-Sham density functional theory (KSDFT) and path integral quantum Monte Carlo (PIMC) \cite{Dornheim_review}. 
In this work, we focus on time-dependent Kohn-Sham density functional theory (TDDFT)~\cite{marques2012fundamentals} to calculate the DSF of electrons from first principles.

Two flavors of TDDFT are utilized to calculate the DSF in the WDM regime: one approach is real-time propagation of Kohn-Sham orbitals (RT-TDDFT) \cite{Baczewski_prl_2016, white2024dynamicalstructurefactorswarm}, while the other employs linear-response theory formulated using Kohn-Sham orbitals (LR-TDDFT) \cite{Mo_prl_2018, Mo_prb_2020, Moldabekov_PRR_2023,Schoerner_PRE_2023}. 

In partially ionized WDM, the number of required orbitals in the KSDFT simulations increases rapidly with temperature. In this context, RT-TDDFT is generally considered more computationally favorable than LR-TDDFT for systems with a large number of thermally occupied orbitals \cite{Dornheim_review}. 
This observation holds true when comparing the conventional LR-TDDFT method, which directly solves Dyson's equation using the orbital representation. This method requires a significant number of unoccupied bands in addition to the fully and partially occupied orbitals of the equilibrium (ground) state \cite{LRT_GPAW2}. We refer to this approach as the \textit{standard} LR-TDDFT. 
In the standard LR-TDDFT method, solving Dyson's equation involves time-consuming inversions and multiplications of large matrices for each frequency value, which is also highly memory-intensive. Consequently, this limits the standard LR-TDDFT method in the WDM regime to relatively small systems, typically in the order of ten atoms in the simulation cell.

An alternative approach for computing the DSF is based on density-matrix representations of LR-TDDFT  using the Liouville–Lanczos (LL) approach, which is also known as time-dependent density-functional perturbation theory \cite{Walker_PRL_2006, Rocca_JCP_2008, Osman_cpc_2011}. This method circumvents the expensive matrix inversions and multiplications associated with standard LR-TDDFT \cite{TIMROV2015460}. Moreover, unlike standard LR-TDDFT, the LL approach does not require additional empty (virtual) bands for calculating the DSF and can access a wide energy (frequency) range, which includes the core-loss region (produced by exciting inner-shell electrons) defined by the utilized pseudopotential. The LL approach for modeling the dynamic density response function of electrons in solids has been implemented as a component in Quantum ESPRESSO (QE) under the name turboEELS \cite{TIMROV2015460}, which uses a recursive Lanczos algorithm demonstrating efficient scalability and parallelization over k-points, band groups, plane-wave schemes, and which allows for restarting from previously interrupted calculations \cite{Carnimeo_JCTC_2023}.
 The LL approach for computing the electronic energy loss spectrum (EELS) at ambient conditions has shown good agreement with standard LR-TDDFT results derived from solving Dyson's equation for gold \cite{Motornyi_prb_2020}, bismuth \cite{Timrov_prb_2017}, carbon \cite{TIMROV2015460}, silicon and aluminum \cite{Timrov_prb_2013}.

Despite its advantages, the LL method has been largely overlooked in the WDM community and has not been utilized to its full potential. To fully exploit the capabilities of the LL method for high-accuracy calculations of the DSF, as well as other related density response and dielectric properties in the WDM regime, the first step is to benchmark and validate the LL approach at elevated temperatures and high densities.

In this work, we benchmark the DSF computed using the LL method against exact PIMC reference results~\cite{Dornheim_MRE_2024} and standard LR-TDDFT at typical warm dense matter conditions. First, we consider isochorically heated electrons in aluminum (Al) with a face-centered cubic (fcc) lattice. This example is relevant for X-ray-driven heating experiments, characterized by heated electrons in a cold ionic lattice ~\cite{Ng1995,Matthieu2011,White2014, Moldabekov_prr_2024}. For isochorically heated Al, we test the LL method against the standard LR-TDDFT within the projector augmented-wave (PAW) formalism~\cite{BlochlPAW} by considering DSFs at different wavenumbers and by considering various types of pseudopotentials. The second test case examines partially degenerate equilibrium warm dense hydrogen at metallic and solid hydrogen densities. These types of warm dense hydrogen are relevant for ICF and hydrogen jet experiments \cite{Fletcher_2022, Zastrau_prl_2014}. 
Since computing dynamic properties from PIMC simulations is rather difficult~\cite{JARRELL1996133,dornheim_dynamic}, we benchmark the LL approach using exact PIMC data for the imaginary-time density--density correlation function (ITCF) of warm dense hydrogen \cite{Dornheim_MRE_2024}. We also provide an analysis of the effect of Lorentzian smearing used in the LR-TDDFT calculations of ITCF. Finally, as an example of the utility of the LL approach, we examine finite size effects in the simulation of the DSF of the warm dense hydrogen at considered densities.

The paper is organized as follows: In Section \ref{s:methods}, we present a brief discussion of the theoretical methods and outline the simulation details. The results of the simulations and their analysis are discussed in Section \ref{s:results}. Finally, we conclude the paper by summarizing our findings and providing an outlook.

\section{Methods and Simulation Details}\label{s:methods}
\subsection{Dynamical structure factor and imaginary-time density--density correlation function}

The primary motivation for calculating the DSF of WDM are XRTS measurements, which constitute a key method of diagnostics for extreme states of matter~\cite{siegfried_review,Gregori2008,kraus_xrts}. In these experiments, the XRTS signal is a convolution of the combined source-and-instrument function $ R(\omega)$ with the DSF of electrons. Specifically, the function $R(\omega)$ takes into account broadening from the detector setup and the characteristics of the X-ray source~\cite{Gawne_JAP_2024}.

Without a loss of generality, the electronic DSF can be broken down into two components: a quasi-elastic contribution from bound electrons and the screening cloud surrounding ions~\cite{Vorberger_PRE_2015}, and an inelastic contribution that arises from electrons transitioning between available states. In the Chihara approach, these are distinguished into those various combinations of transitions between free and bound states~\cite{boehme2023evidence, Chihara_1987, Chihara_JoP_2000}.
The TDDFT method and KS-DFT-based molecular dynamics (MD) of ions allow one to obtain the total electronic DSF, including both the quasi-elastic and inelastic features~\cite{Mo_prb_2020}, but unlike the Chihara model there is no need to distinguish bound and free electrons since all KS states are treated on equal footing. In practice, this is done by two separate sets of simulations. The quasi-elastic component of the DSF is calculated using the Rayleigh weight \cite{Vorberger_PRE_2015,dornheim2024_Rayleigh}, which is determined by the static structure factors for electron-ion and ion-ion pairs from KS-DFT MD. The inelastic part of the DSF is then calculated using the TDDFT method.
Using this approach in WDM conditions is warranted due to the significant differences in the characteristic times scales of electron and ion dynamics. In this work, we limit ourselves to the inelastic part of the DSF from LR-TDDFT. 

In LR-TDDFT, the DSF is computed using the electronic dynamic density response function $\chi(\vec q, \omega)$ in the fluctuation-dissipation theorem \cite{quantum_theory}: 
\begin{equation}
    S(\vec q, \omega)=-\frac{\hbar^2}{n}~\frac{1}{1-e^{-\hbar \omega/k_BT}}~{\rm Im}\left[\chi(\vec q, \omega)\right],
\end{equation}
where $\omega$ is the frequency (or the energy loss in XRTS), $n$ is the electronic density, $T$ is the (electronic) temperature, $\hbar$ is the reduced Planck constant, and $k_B$ is the Boltzmann constant.
In the linear response theory of extended systems with periodic boundary conditions, $\chi(\vec q, \omega)$ is defined as the diagonal part of the microscopic density response function $\chi_{\scriptscriptstyle \mathbf G \mathbf G}(\mathbf k, \omega)$, where $\vec q=\vec G+ \vec k$ and $\vec G$ is a reciprocal lattice vector  (with  $\vec k$ being restricted to the first Brillouin zone) \cite{DSF_LR-TDDFT, martin_reining_ceperley_2016}. In the standard LR-TDDFT,  $\chi_{\scriptscriptstyle \mathbf G \mathbf G}(\mathbf k, \omega)$ is given by the solution of Dyson’s equation  \cite{Byun_2020, martin_reining_ceperley_2016}:
\begin{equation}\label{eq:Dyson}
\begin{split}
\chi_{\scriptscriptstyle \mathbf G \mathbf G^{\prime}}(\mathbf k, \omega)
&= \chi^0_{\scriptscriptstyle \mathbf G \mathbf G^{\prime}}(\mathbf k, \omega)+ \displaystyle\smashoperator{\sum_{\scriptscriptstyle \mathbf G_1 \mathbf G_2}} \chi^0_{\scriptscriptstyle \mathbf G \mathbf G_1}(\mathbf k, \omega) \big[ v_{\scriptscriptstyle \mathbf G1}(\vec k)\delta_{\scriptscriptstyle \mathbf G_1 \mathbf G_2} \\
&+ K^{\rm xc}_{\scriptscriptstyle \mathbf G_1 \mathbf G_2}(\mathbf k, \omega) \big]\chi_{\scriptscriptstyle \mathbf G_2 \mathbf G^{\prime}}(\mathbf k, \omega),
\end{split}
\end{equation}
where $\chi^{~0}_{\scriptscriptstyle \vec G,\vec G^{\prime}}(\vec k,\omega)$ denotes the Kohn-Sham (KS)  response function  \cite{Hybertsen}, $v_{\scriptscriptstyle \mathbf G1}(\vec k)={4\pi}/{|\mathbf k+\mathbf G_1|^2}$, and $ K^{\rm xc}_{\scriptscriptstyle \vec G_1,\vec G_2}(\vec k, \omega)$ is the exchange-correlation (XC) kernel \cite{Gross_PRL1985}.

The main computational burden of LR-TDDFT lies in solving Dyson’s equation for the density response function. The standard way is to compute $\chi^0_{\scriptscriptstyle \mathbf G \mathbf G^{\prime}}(\mathbf k, \omega)$ and use matrix inversion to solve Dyson’s equation (\ref{eq:Dyson}). An alternative approach is the LL method, which finds the solution by employing an iterative approach derived from time-dependent density-functional perturbation theory using the quantum Liouville equation for the reduced one-electron KS density matrix (see Refs. \cite{Rocca_JCP_2008, TIMROV2015460, GE20142080}). 

The LL approach to LR-TDDFT utilizes the definition of the density response function in terms of the reduced one-electron KS density matrix \cite{TIMROV2015460}:
\begin{equation}\label{eq:chi_LL}
    \chi(\vec q, \omega)={\rm Tr}\Big(\hat n_{\vec q}(\omega)\delta \hat \rho_{\vec q}(\omega)\Big),
\end{equation}
where  $\delta \hat \rho_{\vec q}(\omega)=\hat \rho_{\vec q}(\omega)-\hat \rho^{0}$ denotes the perturbation in the KS density matrix due to a weak perturbation $\delta v_{\rm ext, \vec q}\sim e^{i\vec q \vec r}$ applied to an equilibrium state (defined by the unperturbed density matrix $\rho^{0}(\vec r, \vec r^{\prime})$ and the KS Hamiltonian $\hat H^{0}_{\rm KS}$), and $\hat n_{\vec q}$ denotes the response of the
$\vec q$ Fourier component of the charge-density operator.

To compute $\chi(\vec q, \omega)$, instead of the time-dependent KS equation, a quantum Liouville equation for the reduced one-electron KS density matrix is considered in the LL method:
\begin{equation}\label{eq:rho_L}
    i \frac{\mathrm{d}\hat{\rho}(t)}{\mathrm{d}t}=\left[\hat H_{\rm KS} (t), \hat{\rho}(t)\right],
\end{equation}
where the KS Hamiltonian $\hat H_{\rm KS}=-\frac{\nabla^2}{2}+v_{\rm ext}(\vec r,t)+v_{\rm HXC}(\vec r,t)$ contains the total external potential, $v_{\rm ext}(\vec r,t)$, due to the ions and any additional perturbations,  and $v_{\rm HXC}(\vec r,t)$ corresponding to the Hartree and exchange-correlation contributions. The KS density matrix is represented by the KS orbitals $\phi_i$ and corresponding occupation numbers $f_i$ as $\rho(\vec r, \vec r^{\prime}, t)=\sum_i f_i \phi_i(\vec r, t)\phi_i(\vec r^{\prime}, t)$.    

Considering a weak perturbation $\delta v_{\rm ext}(\vec q, \omega)$,  linearization of Eq. (\ref{eq:rho_L}) and subsequent Fourier transformation allows one to find:
\begin{equation}\label{eq:rho1_LL}
    \left( \omega - \hat L \right)\cdot \delta \hat \rho_{\vec q}(\omega)=\left[\delta v_{\rm ext, \vec q}( \omega), \hat \rho^{0}\right],
\end{equation}
where the Liouvillian superoperator  $\hat L$ is defined as
\begin{equation}
    \hat L \cdot \delta \hat \rho_{\vec q}(\omega)=\left[\hat H^{0}_{\rm KS}, \delta \hat \rho_{\vec q}(\omega)\right]+\left[\delta v_{\rm HXC, \vec q}( \omega), \hat \rho^{0}\right]. 
\end{equation}

From Eq. (\ref{eq:chi_LL}), we see that $\chi(\vec q, \omega)$ can be computed if one has the solution of  Eq. (\ref{eq:rho1_LL}), which can be written as
\begin{equation}
    \delta \hat \rho_{\vec q}(\omega)=\left( \omega-\hat L\right)^{-1}\cdot \left[\delta v_{\rm ext, \vec q}( \omega), \hat \rho^{0}\right] 
\end{equation}

The LL approach is an iterative method computing  a tridiagonal form of the matrix $L\simeq L^{(M)}$ (defining $\hat L$ in a given representation) in terms of the Lanczos coefficients (wit $M$ being the number of  Lanczos iterations). More details of the Lanczos algorithm can be found, e.g., in Ref. \cite{Saad_book}, and specifics of the implementation for TDDFT (the Liouville-Lanczos method) are presented in
Refs. \cite{Rocca_JCP_2008, Osman_cpc_2011, GE20142080, TIMROV2015460, Timrov_prb_2013}. 

In Dyson's equation (\ref{eq:Dyson}), calculating the response function $\chi^{~0}_{\scriptscriptstyle \vec G,\vec G^{\prime}}(\vec k,\omega)$ requires matrix elements that represent transitions between states with a momentum difference significantly larger than \(\vec q\) and an energy eigenvalue difference that is considerably greater than \(\omega\) \cite{LRT_GPAW2}. This necessity means that a substantial number of additional bands are needed beyond those required to compute the equilibrium (unperturbed) density matrix $\rho^{0}$. In contrast, the Lanczos algorithm used in the LL method only relies on information in $\rho^{0}$ and does not require any extra empty bands.

Theoretically, the standard LR-TDDFT method solving Dyson’s equation and LL approach should yield the same solution for $\chi(\vec{q}, \omega)$. However, variations in implementation details---such as the choice of basis sets for wavefunctions, handling of tightly bound electrons, and the numerical solvers used---can introduce numerical differences that lead to deviations between the results. As far as we are aware, the LL method for $\chi(\vec q, \omega)$ has only been implemented in Quantum ESPRESSO \cite{TIMROV2015460}, which utilizes pseudopotential-based KS-DFT with normconserving and ultrasoft pseudopotentials. This implementation is focused on the applications at ambient conditions to model the energy loss spectrum of electrons in solids.
To test this implementation of the LL method at WDM conditions, we compare its results for the DSF with the data computed using the standard LR-TDDFT based on the PAW formulation. Furthermore, we validate the LL approach by examining the ITCF results and comparing them with the data from PIMC simulations for warm dense hydrogen.

The connection between the DSF and the ITCF reads:
\begin{eqnarray}\label{eq:analytic_continuation}
F(\mathbf{q},\tau) = \int_{-\infty}^\infty \textnormal{d}\omega\ S(\mathbf{q},\omega)\ e^{-\tau\omega},
\end{eqnarray}
with $-i\hbar\tau\in -i\hbar[0,\beta]$ being the imaginary time argument in the range defined by the inverse temperature $\beta=1/(k_BT)$. 
For the calculation of $F(\mathbf{q},\tau)$, the detailed balance $S(\mathbf{q},-\omega)=S(\mathbf{q},\omega)\exp(-\beta \omega)$ is used to compute the DSF at negative frequencies. 
High-accuracy data for the ITCF can be obtained from various types of quantum Monte Carlo simulations~\cite{Vitali_PRB_2010,Filinov_PRA_2012,Boninsegni_maximum_entropy,dornheim_dynamic,Ferre_PRB_2016,Motta_JCP_2015,Filinov_PRA_2016,Dornheim_SciRep_2022,Dornheim_insight_2022}. The one-to-one connection (\ref{eq:analytic_continuation}) between the DSF and the ITCF means that both contain the same information \cite{Dornheim_prb_2023, Dornheim_insight_2022}. 
However, inverting Eq. (\ref{eq:analytic_continuation}) to obtain the DSF from the ITCF poses a numerically challenging problem \cite{chuna_paper1_2025}. This inversion is ill-posed concerning the Monte Carlo error bars of $ F(\mathbf{q},\tau) $ and is associated with potential numerical instabilities \cite{JARRELL1996133,Goulko_PRB_2017}. 
Consequently, instead of inverting Eq. (\ref{eq:analytic_continuation}), we compute the ITCF from the DSF results obtained using the LL method and subsequently compare it with the ITCF data acquired from PIMC simulations.

As mentioned above, LR-TDDFT does not capture the quasi-elastic part of the DSF, which translates to a constant shift of the ITCF that does not depend on $\tau$. To circumvent this aspect, we compare the shifted  ITCF values:
\begin{equation}
    \widetilde F(\vec q, \tau)=F(\vec q, \tau)-F(\vec q, \tau=0),
    \label{eq:tilde_ITCF}
\end{equation}
which still allows for testing all dynamic features translated to the imaginary time domain \cite{Dornheim_prb_2023, Dornheim_insight_2022, Dornheim_PTR_2022}.

In practice, the Lorentzian smearing parameter $\eta$ is introduced into the LR-TDDFT calculations to regularize $\chi(\vec q, \omega)$ (both in the standard LR-TDDFT and the LL method). This is done by adding a small imaginary part to the frequency, transforming it from $\omega$ to $\omega + i\eta$. As a result, discrete spectral features (lines) in the DSF are broadened \cite{Rocca_JCP_2008}. In the real frequency domain, this technique serves as a method to compute a continuous $\chi(\vec q, \omega)$, with a negligible impact on the shape and features of the DSF $S(\vec q, \omega)$. However, the effect of Lorentzian smearing on the ITCF $F(\vec q, \tau)$, computed using $S(\vec q, \omega)$ from LR-TDDFT, has not been considered in previous studies. This is a relevant question because XRTS measurements can be analyzed in the Laplace domain, which helps to circumvent the deconvolution problem of the XRTS spectrum with the source and instrument function $R(\omega)$ \cite{Dornheim_insight_2022, Dornheim_T2_2022}. Therefore, we also examine the impact of Lorentzian smearing employed in LR-TDDFT on the ITCF.
We note that the broadening parameter $\eta$ is a feature of LR-TDDFT and is not needed in the PIMC simulations of the ITCF.
 
%%%%%%%%%%%%%%%%%%%%%%%%%%%%%%%%%%%%%%%%%%%%%%%%%%%%%%%%%%%%%%%%%%%%%%%%%%%%%%%%%%%%%%%%%%%%%%%%%% 
\subsection{Simulation details}

The standard LR-TDDFT calculations based on the solution of Dyson’s equation (\ref{eq:Dyson}) within the PAW method have been performed using the GPAW code~\cite{GPAW1, GPAW2, LRT_GPAW1, LRT_GPAW2, ase-paper, ase-paper2}, which is a real-space implementation of the PAW method~\cite{BlochlPAW}.
For the LL approach to compute DSF, we used Quantum ESPRESSO \cite{Giannozzi_2009, Giannozzi_2017, Giannozzi_jcp_2020, Carnimeo_JCTC_2023, TIMROV2015460}. A static (adiabatic) LDA approximation  $K^{\rm xc}_{\scriptscriptstyle \vec G_1,\vec G_2}(\vec k, \omega=0)$ for the XC kernel was used in all calculations.

For Al, considering the [100] crystallographic direction, we used a $20\times20\times20$ $k$-point grid and an energy cutoff of $500~{\rm eV}$. A conventional unit cell with a lattice parameter of the fcc crystal $a=4.05 ~\textup{\AA}$ was considered~\cite{wyckoff1948crystal}. 
The electron temperature in Al was set to $T=6~{\rm eV}$.
The results were computed using the Lorentzian smearing parameter  $\eta=0.5~{\rm eV}$, which is small enough to distinguish differences between the LL method and the standard LR-TDDFT. 
The local density approximation (LDA) and generalized gradient approximation (GGA) level XC functionals were considered.
We used $N_b=200$ KS bands. In GPAW calculations, we utilized the LDA and PBE PAW setups provided by GPAW, which were generated using the Perdew-Wang \cite{PhysRevB.45.13244} and PBE \cite{PhysRevLett.77.3865} functionals, respectively.
For the LL calculations, the Perdew-Zunger LDA functional \cite{Perdew_Zunger_PRB_1981} was used with the  \textsc{Al.pz-vbc.UPF} pseudopotential, the GGA level PBE functional was used in combination with normconserving pseudopotential \textsc{Al.pbe-rrkj.UPF}. Both pseudopotentials are from the Quantum ESPRESSO pseudopotential database \cite{PP_ESPRESSO_2022}.

\begin{figure}
    \centering
    \includegraphics[width=0.48\textwidth,keepaspectratio]{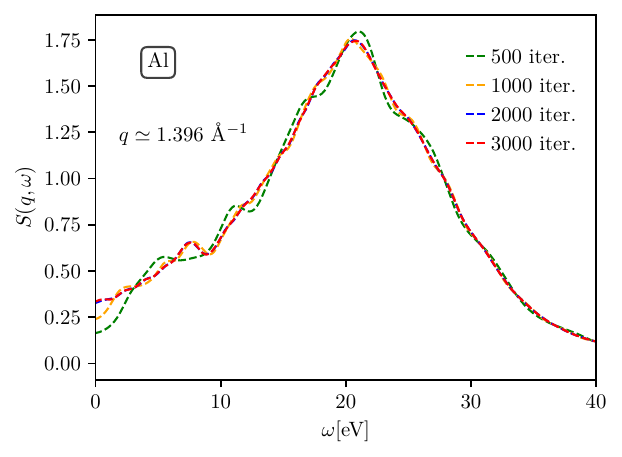}
    \caption{ Convergence with respect to the number of Lanczos iterations in LL calculations of the electronic DSF of isochorically heated Al at $T=6~{\rm eV}$.
    The LL calculation results with 2000 Lanczos iterations (blue curve) are visually nearly indistinguishable from the results with 3000 Lanczos iterations (red curve). 
    }
    \label{fig:Al_iter}
\end{figure}

For hydrogen, we tested various types of pseudopotentials in combination with the PBE XC functional. 
In the calculations using Quantum ESPRESSO, the used pseudopotentials are normconserving \textsc{H.pbe-vbc.UPF} \cite{PP_ESPRESSO_2022} and ultrasoft $\textsc{h\_pbe\_v1.4.uspp.F.UPF}$ \cite{Garrity_CMS_2014}.
In addition, the PAW method was used with the PBE setup for hydrogen from GPAW. We considered $14$, $32$, and $100$ protons in the simulation cell. As a part of the analysis of the behavior of the ITCF computed from the DSF values, we varied the Lorentzian smearing parameter in the range from $\eta=0.2~{\rm eV}$ to $\eta=3.0~{\rm eV}$.
The DSFs were computed for warm dense hydrogen in equilibrium at densities $\rho=0.08 ~{\rm g/cc}$ ($T=4.8~{\rm eV}$) and $\rho=0.34~{\rm g/cc}$ ($T=12.528~{\rm eV}$). These correspond to the density parameters $r_s=3.23$ and  $r_s=2.0$, with the parameter $r_s$ being the ratio of the mean interparticle distance to the Bohr radius (Wigner-Seitz radius)~\cite{Ott2018}. The considered temperatures provide a partially degenerate state with the system temperature $T$ equal to the Fermi temperature of electrons $T_F$. We choose these densities and temperatures to benchmark TDDFT results against the available PIMC results for the ITCF \cite{Dornheim_MRE_2024}. 
The size of the simulation box is determined by the density $r_s$ (or $\rho$) and the number of protons $N$, expressed as $L = r_s \left( \frac{4\pi}{3}N \right)^{1/3}$ (in atomic units). We set the Monkhorst-Pack $k$-point grid to $10\times10\times10$ for $N=14$, $8\times8\times8$ for $N=32$, and $2\times2\times2$ for $N=100$. The DSF results were averaged over 10 snapshots for $N=14$ and $N=32$ at $r_s=2$. For $N=100$ at $r_s=2$, the DSF values were averaged over 5 snapshots. At $r_s=3.23$, we computed the DSF values using $N=14$ and $N=32$ protons with averaging performed over 20 snapshots. The snapshots of the position of protons were chosen randomly from the particles trajectory obtained by the KSDFT-based MD simulations.
In all calculations of the DSF, we used a bi-constant extrapolation scheme after the computed number of LL iteration $N_{\rm iter}$ to obtain  $10^4$ Lanczos coefficients \cite{TIMROV2015460}. For Al, we used $N_{\rm iter}=3\times10^3$ and for warm dense hydrogen $N_{\rm iter}=1.2\times 10^4$.

%%%%%%%%%%%%%%%%%%%%%%%%%%%%%%%%%%%%%%%%%%%%%%%%%%%%%%%%%%%%%%%%%%%%%%%%%%%%%%%%%%%%%%%%%%%%%%%%%%
\section{Results} \label{s:results}
\subsection{Isochorically heated aluminum}
We begin by analyzing the DSF of isochorically heated Al, which has cold ions arranged in the fcc lattice structure and hot electrons at a temperature of $T = 6~{\rm eV}$. Such a transient state is generated and examined in experiments that utilize X-ray-driven heating used for XRTS measurements \cite{ Descamps_sciadv, Descamps2020, Matthieu2011, White2014}. In these experiments, the incident X-rays do not directly interact with the nuclei, keeping them cold during a free electron laser pulse that lasts less than 100 femtoseconds.

The simulations were performed for wavenumbers $q=0.155~{\rm \AA^{-1}}$, $q=0.7757~{\rm \AA^{-1}}$, and $q=1.396~{\rm \AA^{-1}}$ along the crystallographic direction [100]. 
In the standard LR-TDDFT method, the wavenumbers must correspond to the difference between two $k$-points. 
The LL method does not have this constraint.
We note that more flexibility in choosing $\vec q$ values in the LL method provides an additional advantage for the analysis of the experimental XRTS data where the wavenumber blurring effect occurs due to the finite size of a detector \cite{Gawne_prb_2024}. To compare the results from Quantum ESPRESSO and GPAW, we select the wavenumber values in the LL calculations to correspond with those derived from the differences between two $k$-points at the specified parameters.

\begin{figure}
    \centering
    \includegraphics[width=0.48\textwidth,keepaspectratio]{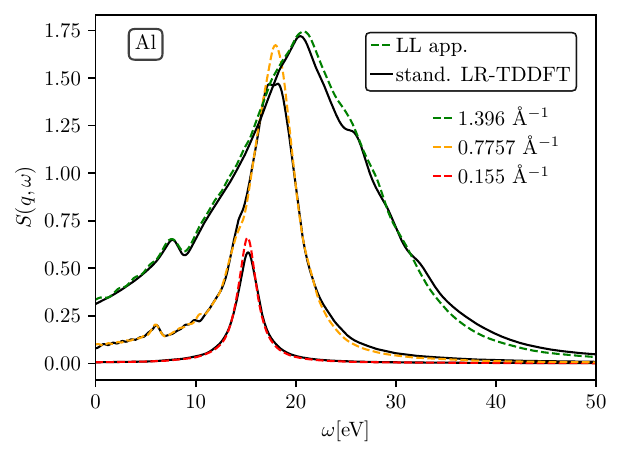}
    \caption{Comparison of the DSF of isochorically heated Al computed using the LL method and the standard LR-TDDFT at different momentum transfer values for $T=6~{\rm eV}$.}
    \label{fig:Al_qe_vs_gpaw_lda}
\end{figure}

\begin{figure}
    \centering
    \includegraphics[width=0.48\textwidth,keepaspectratio]{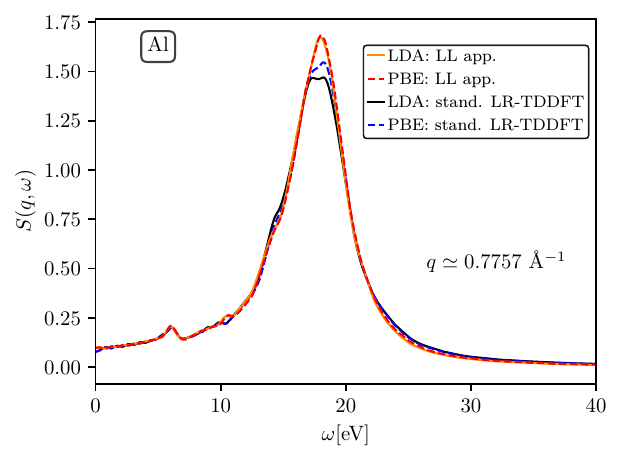}
    \caption{DSF results for isochorically heated Al at $T=6~{\rm eV}$ using LDA and GGA level XC functionals in the LL method employing norm-conserving pseudopotentials and the standard LR-TDDFT within the PAW framework.}
    \label{fig:Al_LDA_vs_PBE}
\end{figure}

\begin{figure}\centering
\includegraphics[width=0.3\textwidth]{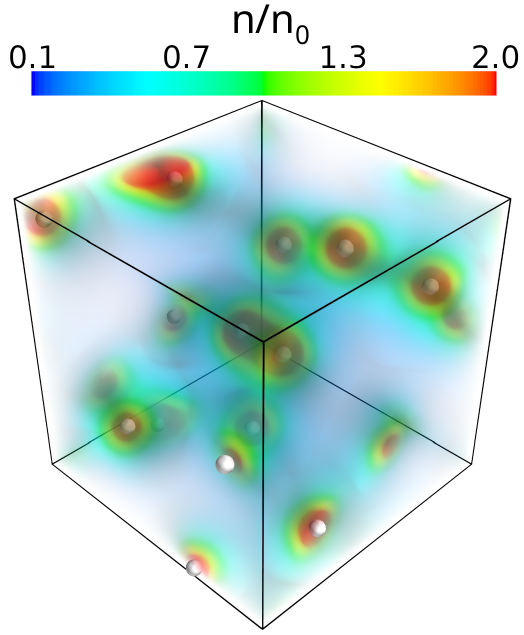}
\includegraphics[width=0.48\textwidth]{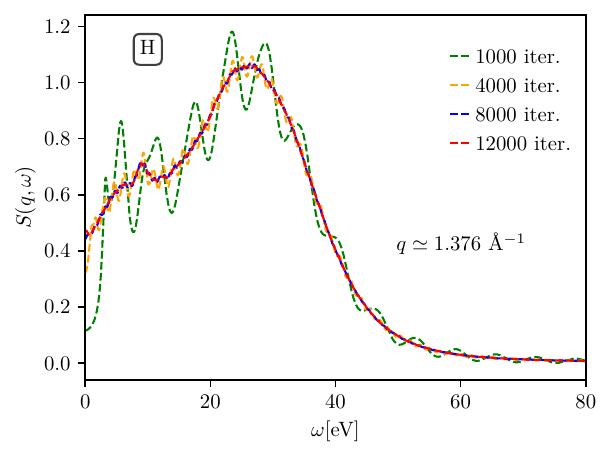}
\includegraphics[width=0.48\textwidth]{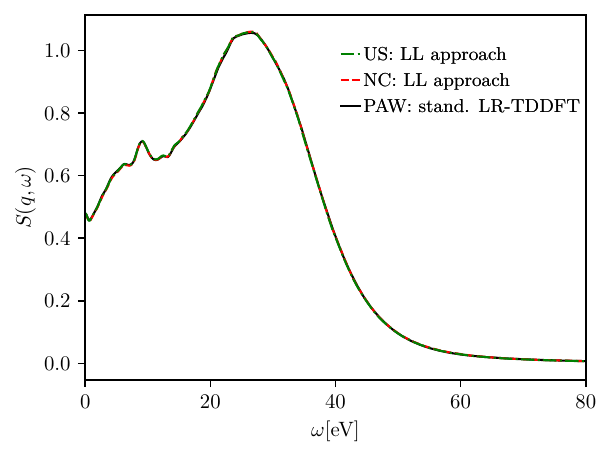}
\caption{\label{fig:H_N14_snp} 
The top panel shows the density distribution in the simulation cell (in units of mean density $n_0$) with $N=14$ protons at $r_s=2$ and $T=12.58~{\rm eV}$. The middle panel illustrates the convergence of the DSF calculations of warm dense hydrogen with respect to the number of Lanczos iterations in the LL method.
The bottom panel compares the DSF results computed using ultrasoft (denoted as US) and norm-conserving (denoted as NC) pseudopotentials in the LL method with standard LR-TDDFT within the PAW framework.  
}
\end{figure}

In Fig. \ref{fig:Al_iter}, we show the convergence of the results for isochorically heated Al with respect to the number of iterations $N_{\rm iter}$ in the LL calculations at $q=1.396~{\rm \AA^{-1}}$. From Fig. \ref{fig:Al_iter}, it is evident that the LL results are converged after $N_{\rm iter}=2000$ iterations. In Fig. \ref{fig:Al_qe_vs_gpaw_lda}, considering the LDA level description of the XC functional, the LL results computed using $N_{\rm iter}=3000$ are compared with the exact solution of  Dyson's equation within the PAW formalism. Overall, we find good agreement between the different approaches.
%agreement between data calculated using considered different approaches is good. 
At $q=0.155~{\rm \AA^{-1}}$, one can see a well-defined plasmon feature, which becomes broader with the increase in the wavenumber, indicating a stronger Landau damping effect~\cite{Lan67}. The pair continuum (strong plasmon damping) for Al corresponds to $q>1.3~{\rm \AA^{-1}}$, where the plasmon is not well defined anymore \cite{quantum_theory, Hamann_cpp}. A small spike at $\omega<10~{\rm eV}$ at $q=0.7757~{\rm \AA^{-1}}$ and $q=1.396~{\rm \AA^{-1}}$ is due to lattice effects leading to energy gaps at Brillouin zone boundaries \cite{Larson_2000}.
An interesting observation from Fig. \ref{fig:Al_qe_vs_gpaw_lda} is a noticeable difference in the magnitude of the DSF maximum for $q=0.7757~{\rm \AA^{-1}}$ between the LL method and Dyson's equation solution within the PAW formalism. To analyze this aspect further, we compare the results of the DSF computed using the PBE XC functional with those obtained using LDA level functionals, as shown in Fig. \ref{fig:Al_LDA_vs_PBE}. The calculations were performed using norm conserving pseudopotentials and PAW setups that are consistent with the XC functionals used. The PBE-based results from the LL method are in close agreement with the LDA-level calculations using the same method.
When examining the DSFs obtained from solving Dyson's equation within the PAW framework, we find that the maximum of the DSFs calculated with the PBE functional are larger than those derived from the LDA-level data. This larger value leads to closer agreement with the results obtained from the LL method. We conclude that the observed differences in the vicinity of the DSF maximum are due to the differing methods (pseudopotentials) used to handle core electrons.

\begin{figure}
    \centering
    \includegraphics[width=0.48\textwidth,keepaspectratio]{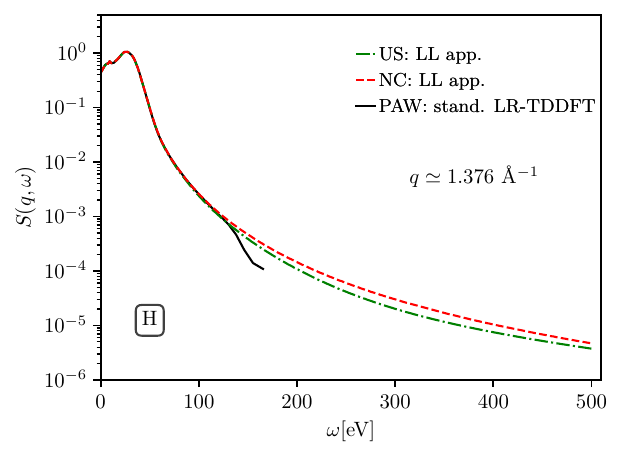}
    \caption{The same as in the bottom panel of Fig. \ref{fig:H_N14_snp}, but presented over  wider frequency range $\omega$ using a logarithmic scale.}
    \label{fig:H_14_log}
\end{figure}

\subsection{Warm dense hydrogen at metallic density}\label{sub:rs2}

As an example of WDM in an equilibrium state, we examine warm dense hydrogen.
We start with a metallic density with $r_s=2$ and at degeneracy parameter $\theta=T/T_F=1$. This is a typical WDM regime where neither correlations nor thermal excitations nor quantum effects can be neglected.
In Fig. \ref{fig:H_N14_snp}, we show results for one snapshot of 14 protons at $q=1.376~{\rm \AA^{-1}}$. 
The top panel of Fig. \ref{fig:H_N14_snp} shows the considered snapshot along with the density distribution of electrons $n$, expressed in terms of the mean electron density $n_0$. From this panel, we can observe the characteristic electron structure in WDM, where the distinction between electrons that are localized around ions and those that are quasi-free is not clearly defined \cite{Moldabekov_jctc_2024}. In the middle panel of Fig. \ref{fig:H_N14_snp}, we illustrate the convergence of the results with respect to the number of iterations $N_{\rm iter}$ in the LL calculations at $q=1.376~{\rm \AA^{-1}}$ and smearing parameter $\eta=0.5~{\rm eV}$, where one can see that the results are converged after $N_{\rm iter}=8000$ iterations.  
In the bottom panel of Fig. \ref{fig:H_N14_snp}, we provide the comparison of the data computed using norm conserving and ultrasoft pseudopotentials in the LL method with the results computed solving Dyson's equation using the PAW approach. 
From this comparison, we see perfect agreement between the different approaches considered at $0 ~{\rm eV}\leq \omega \leq 80~{\rm eV}$. Using a logarithmic scale, in Fig. \ref{fig:H_14_log}, we show the same data in the range $0 ~{\rm eV}\leq \omega \leq 500~{\rm eV}$. From Fig. \ref{fig:H_14_log}, it is visible that the DSF calculated using the standard LR-TDDFT method shows a drop starting around 150 eV. This drop results in a strong deviation from the results obtained using the LL approach. Besides that, the standard LR-TDDFT is limited to 180 eV at the used number of orbitals ($N_b=200$). These features are attributed to the dependence on the number of unoccupied states in the standard LR-TDDFT method, where the maximum value of $\omega$ is limited to the largest difference in the energy of computed states. In contrast, the LL method does not require extra empty states for computing the DSF at large frequencies (energies) $\omega$, as demonstrated in Fig. \ref{fig:H_14_log}. 

%The demonstrated limitation in the frequency range caused by the finite number of empty states in the standard LT-TDDFT method restricts the values of the wavenumber  $q $ for which the DSF can be computed in a relevant range of $\omega$.
The demonstrated limitation in the frequency ranged caused by the finite number of empty states in the standard LT-TDDFT method restricts its use to predicting the DSF to relatively small energy losses. In practice, this means that many interesting features and cases are beyond the reach of the standard LR-TDDFT. For example, all elements beyond boron have at least one excitation edge beyond 200~eV~\cite{ExcitationEnergies,ExcitationEnergies2}. Similarly, examining the complete DSF at high $q$ is inaccessible to standard LR-TDDFT simulations.
The increase in $q$ leads to the shift of the DSF maximum position $\omega_0$ to larger values, which leads to the increase in the number of required empty orbitals in the standard LR-TDDFT. 
In WDM, as well as metals and some non-metallic systems, where electrons exist in or are easily excited into the conduction band, the energy of the DSF maximum, $ \omega_{0}$, shows $\omega_0^2-\omega_p^2\sim q^2$ scaling at small wavenumbers transiting to quadratic dependence $ \omega_{0} \sim q^2 $ at large wavenumbers \cite{Arista_pra_1984, Moldabekov_pop_2018, Gawne_prb_2024, Dornheim_roton_2022}, where $\omega_p$ denotes electron plasma frequency.
To accurately compute the DSF, it is necessary to have a sufficient number of empty bands whose energy differences cover the range $ \varepsilon \gg \omega_0$. By using the density of states for a free electron gas as an approximation, the number of electron orbitals within a specified energy range $ \varepsilon $ can be estimated as $ N_b \sim \varepsilon^{3/2}$, leading to cubic scaling $N_b\sim q^3$ at large wavenumbers. This cubic scaling of the required number of bands results in significant memory demands for parallel computations in the standard LR-TDDFT method. This makes the standard LR-TDDFT method impractical for modeling the DSF at large wave numbers using a meaningful frequency grid and range. 
For example, at very high $q$, where the single-particle limit is being probed, the DSF is characterized by a Compton feature, which can be positioned at frequencies in excess of 100 eV and have a width of more than 100 eV (see e.g. Ref.~\cite{kraus_xrts}), which would require a huge number of bands to model. Simulating the DSF at very high $q$ with standard LR-TDDFT would, therefore, be even more challenging.
Furthermore, modeling the DSF of disordered systems incurs additional computational costs due to the necessity of averaging over multiple snapshots.

\begin{figure}
\centering
\includegraphics[width=0.47\textwidth]{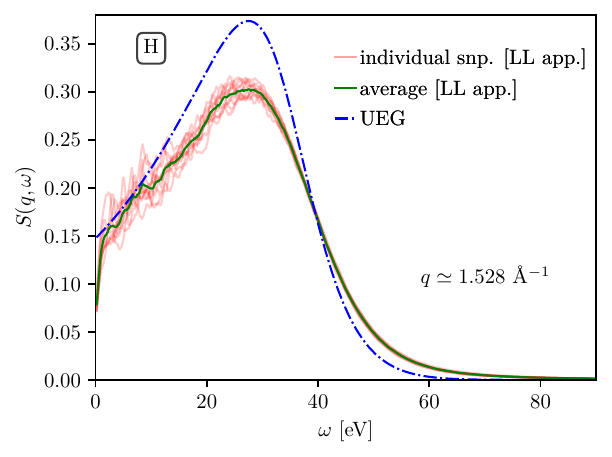}
\includegraphics[width=0.48\textwidth]{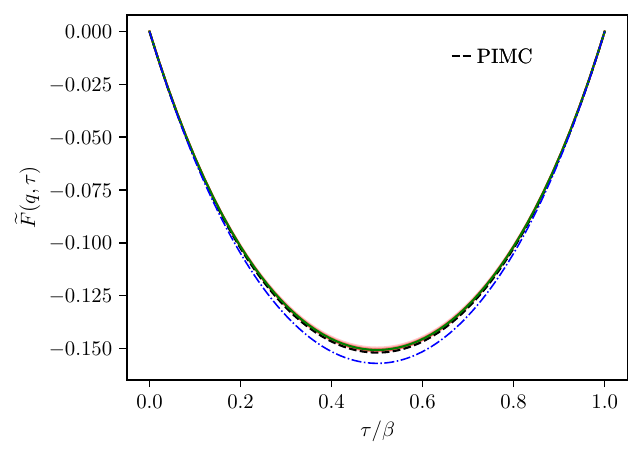}
\caption{\label{fig:H_14_av_qmin}
Top panel: DSF of warm dense hydrogen computed using 10 different snapshots of $N=14$ protons (red lines) and corresponding averaged result (green line). Bottom panel: the shifted ITCF results defined by Eq. (\ref{eq:tilde_ITCF}) from the calculations using the LL method and PIMC. The DSF and ITCF results are compared with the data computed for a UEG model~\cite{Dornheim_PRL_2020_ESA}. The LR-TDDFT  and PIMC results were computed at $r_s=2$, $T=12.58~{\rm eV}$, $q\simeq 1.528~{\rm \AA^{-1}}$, and using the Lorentzian smearing parameter $\eta=0.2~{\rm eV}$. 
}
\end{figure} 

\begin{figure}\centering
\includegraphics[width=0.48\textwidth]{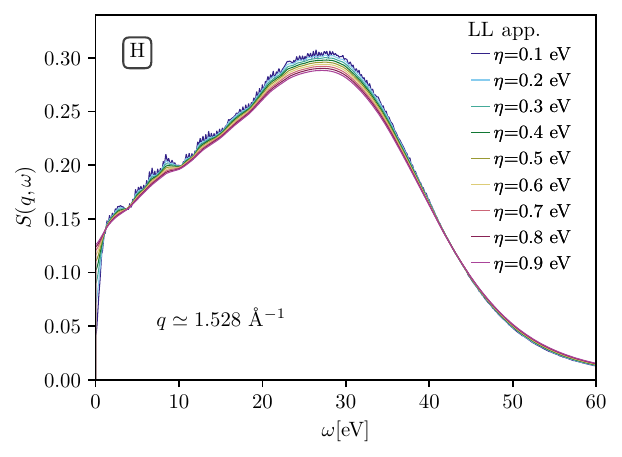}
\includegraphics[width=0.48\textwidth]{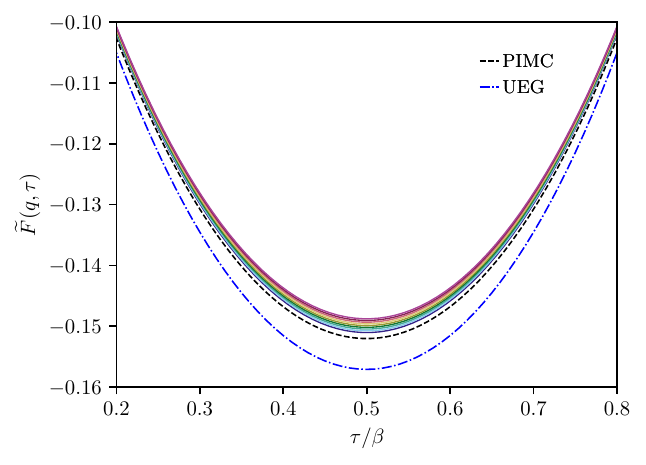}
\caption{\label{fig:eta_H_14_av_qmin} 
Simulation results of the DSF (top panel) and shifted ITCF (bottom panel) for warm dense hydrogen at $r_s=2$ and $T=12.58~{\rm eV}$. The Lorentzian smearing parameter is varied in the range $0.1~{\rm eV}\leq \eta \leq 0.8~{\rm eV}$.
The calculations were performed by averaging over 10 different snapshots of 14 protons. 
}
\end{figure} 

\begin{figure}\centering
\includegraphics[width=0.48\textwidth]{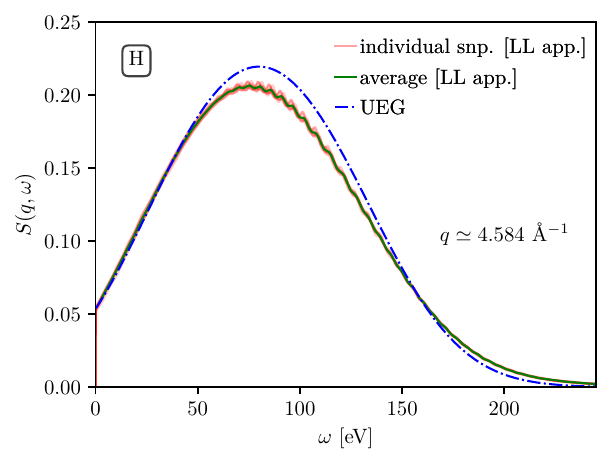}
\includegraphics[width=0.48\textwidth]{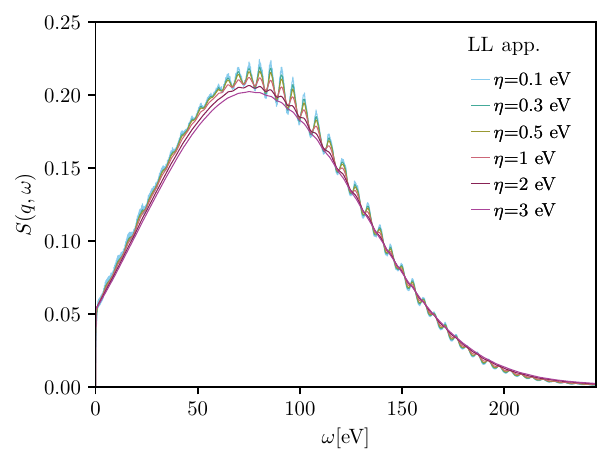}
\includegraphics[width=0.48\textwidth]{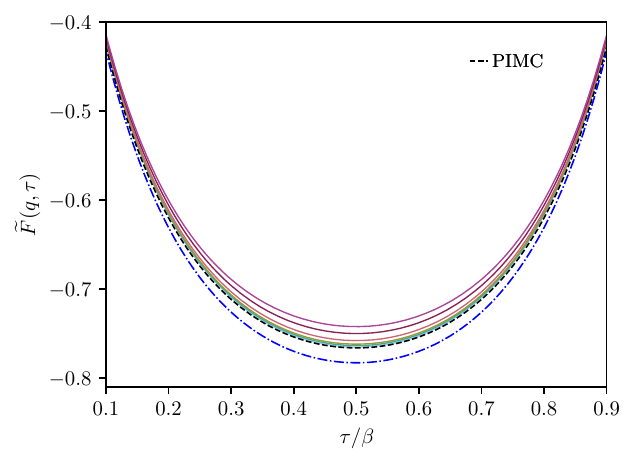}
\caption{\label{fig:eta_H_14_av_3qmin} 
Top panel: DSF of warm dense hydrogen computed using 10 different snapshots of $N=14$ protons (red lines) and corresponding averaged result (green line) at $\eta=2~{\rm eV}$. Middle panel: effect of the variation of the Lorentzian smearing parameter in the range $0.1~{\rm eV}\leq \eta \leq 3.0~{\rm eV}$. 
Bottom panel: shifted ITCF results defined by Eq. (\ref{eq:tilde_ITCF}) from calculations using the LL method (with different $\eta$ values) and PIMC. The DSF and ITCF results are compared with the data computed using the UEG model~\cite{Dornheim_PRL_2020_ESA}. The results are presented for $r_s=2$ and $T=12.58~{\rm eV}$.
}
\end{figure} 

\begin{figure}
    \centering
    \includegraphics[width=0.5\textwidth,keepaspectratio]{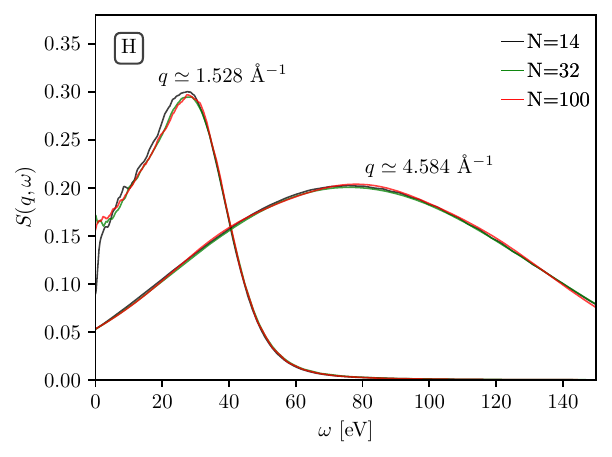}
    \caption{
    Comparison of the DSFs of warm dense hydrogen computed using 14, 32, and 100 particles in the simulation cell at $r_s=2.0$ and $T=12.58~{\rm eV}$. The Lorentzian smearing parameter was set to $\eta=0.3~{\rm eV}$ for $q\simeq 1.528~{\rm \AA^{-1}}$ and to $\eta=3.0~{\rm eV}$ for $q\simeq 4.584~{\rm \AA^{-1}}$. The results were computed using the LL approach to LR-TDDFT. 
    }
    \label{fig:N100}
\end{figure}

An example of averaging over snapshots is illustrated in the top panel of Fig. \ref{fig:H_14_av_qmin} for $q = 1.528~{\rm \AA^{-1}}$ using $ \eta = 0.2~{\rm eV} $. In this panel, we present the DSF values computed from 10 different snapshots, along with the corresponding averaged results. The results for warm dense hydrogen are compared with data from the Uniform Electron Gas (UEG) model, which was computed using an effective static approximation for the local field correction \cite{Dornheim_PRL_2020_ESA,Dornheim_PRB_ESA_2021}.
From the top panel of Fig. \ref{fig:H_14_av_qmin}, we observe that the UEG model significantly overestimates the DSF magnitude at its peak and exhibits a faster decay at frequencies $\omega$ beyond the DSF maximum. 
In the bottom panel of Fig. \ref{fig:H_14_av_qmin}, we compare the shifted ITCF $\widetilde{F}(\vec{q}, \tau)$ (as defined in Eq. (\ref{eq:tilde_ITCF})) calculated using the DSF values obtained from LL-based LR-TDDFT simulations—both averaged and from individual snapshots—against the $\widetilde{F}(\vec{q}, \tau)$   from the PIMC simulations of warm dense hydrogen \cite{Dornheim_MRE_2024} and the data from the PIMC simulation of the UEG. 
This comparison reveals that the averaged LR-TDDFT data for $\widetilde{F}(\vec{q}, \tau)$  aligns excellently with the PIMC results for warm dense hydrogen. In contrast, the UEG model demonstrates substantial differences from the results for warm dense hydrogen, particularly around $\tau=0.5 \beta$.

The shape of the LR-TDDFT results depends on the chosen Lorentzian smearing parameter  $\eta$. A larger $ \eta $ leads to a stronger broadening of the DSF results. To demonstrate that the observed agreement between the LR-TDDFT results (calculated using the LL method) and the PIMC data for $ \widetilde F(q, \tau)$ is not an artifact of the selected $ \eta $ value, we investigate the effects of varying $ \eta $ on $ S(q, \omega) $ and $\widetilde F(q, \tau)$. 
The results for $ S(q, \omega) $ and $ \widetilde F(q, \tau) $, averaged over snapshots, for the range $ 0.1~{\rm eV} \leq \eta \leq 0.9 ~{\rm eV} $ are presented in Fig. \ref{fig:eta_H_14_av_qmin}. From this figure, it is evident that decreasing $ \eta $ leads to noisier results for the DSF, alongside a larger magnitude of the maximum in the DSF. However, for $ \widetilde F(q, \tau) $, a decrease in $ \eta $ results in a closer agreement between the LR-TDDFT results obtained using the LL method and the PIMC data. It is noteworthy that the Laplace transform acts as a noise filter \cite{Dornheim_pop_2023}, resulting in smooth ITCF results even at smaller $ \eta $ values.

As previously mentioned, the LL approach to LR-TDDFT is particularly valuable for DSF calculations at large wavenumbers. To test the application of the LL method for large $ q $ values, we conducted simulations for $ q = 4.584~{\rm \AA^{-1}}$. The results for this wavenumber, at $ r_s = 2 $ and $ \theta = 1 $, are presented in Fig. \ref{fig:eta_H_14_av_3qmin}.
In the top panel, we show results for ten individual snapshots at $\eta = 2~{\rm eV} $, along with the averaged value over these snapshots and data from the UEG model. It is evident that the differences between the DSFs of individual snapshots and the averaged value are much less pronounced compared to the case with $ q = 1.528~{\rm \AA^{-1}} $. Furthermore, at $q = 4.584~{\rm \AA^{-1}}$, the deviation of the UEG model results from the LR-TDDFT data is significantly smaller than that observed at $q = 1.528~{\rm \AA^{-1}}$. Both of these findings are readily explained in terms of the characteristic length scale $\lambda=2\pi/q$. Specifically, the limit of large $q$ corresponds to the structure-less single-particle regime where correlations between electrons and nuclei effectively vanish.

\begin{figure}\centering
\includegraphics[width=0.48\textwidth]{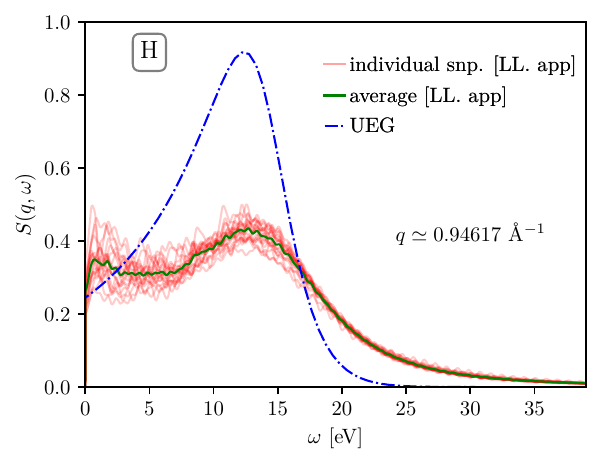}
\includegraphics[width=0.48\textwidth]{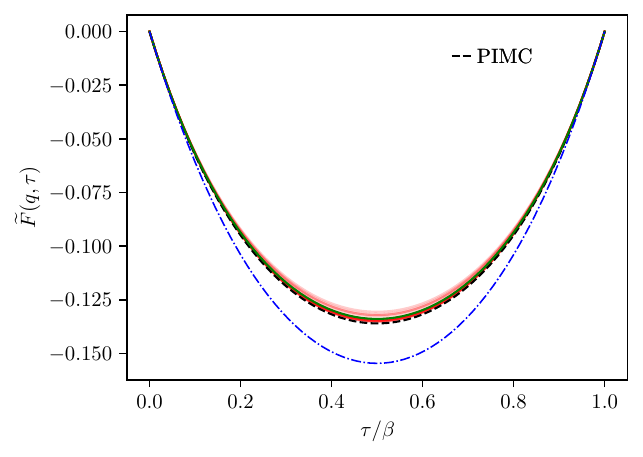}
\caption{\label{fig:UEG_N14_rs2_theta2_switch_probability} 
Simulation results for the DSF (top panel) and shifted ITCF (bottom panel) for warm dense hydrogen at $q\simeq 0.946~{\rm \AA^{-1}}$, $r_s=3.23$ and $T=4.8~{\rm eV}$ with the Lorentzian smearing parameter set to $\eta = 0.3~{\rm eV}$.
The LL method-based LR-TDDFT were performed by averaging over 20 different snapshots of 14 protons. The results are compared with the data computed using the UEG model~\cite{Dornheim_PRL_2020_ESA} and with the PIMC results for warm dense hydrogen \cite{Dornheim_MRE_2024}.
}\label{fig:rs3.23_qmin}
\end{figure} 

In the middle panel of Fig. \ref{fig:eta_H_14_av_3qmin}, we present the results for the snapshot-averaged DSF at $q = 4.584~\text{Å}^{-1}$ for $0.1~\text{eV} \leq \eta \leq 3.0~\text{eV}$. Notably, we observe significantly stronger quasi-periodic noise in the DSF for $\eta < 1~\text{eV}$ compared to that at $q = 1.528~\text{Å}^{-1}$ (see Fig. \ref{fig:eta_H_14_av_qmin}).
A smooth curve for the DSF is produced using $\eta \gtrsim 2~\text{eV}$. 
We note that the magnitude of $\eta$ should be gauged taking into account the characteristic frequency (energy) scale $\omega$ of the DSF.  
For example, the full width at half maximum (FWHM), denoted as $\Delta \omega_0$, at $q = 4.584~\text{Å}^{-1}$ is approximately 3.5 times larger than that at $q = 1.528~\text{Å}^{-1}$. The relative value of $\eta/\Delta \omega$ required for damping the noise in the DSF is similar for both wavenumbers, with an optimal value around $\eta_0 = 10^{-2} \times \Delta \omega_0$. This smearing degree results in a minor effect on the overall shape of the DSF, with the magnitude of the DSF maximum being underestimated by about $5\%$. At $\eta\lesssim \eta_0$, the FWHM $\Delta \omega_0$ and the position of the DSF maxima, $\omega_0$, remain nearly unchanged for both $q = 4.584~\text{Å}^{-1}$ and $q = 1.528~\text{Å}^{-1}$. The increase in $\eta$ gradually deteriorates the DSF maximum prediction due to the DSF asymmetry relative to $\omega_0$.

In the bottom panel of Fig. \ref{fig:eta_H_14_av_3qmin}, we show the shifted ITCF $\widetilde F(q,\tau)$ data computed using the LL method, the PIMC results for warm dense hydrogen, and the ITCF data for UEG from PIMC simulations at $q = 4.584~{\rm \AA^{-1}}$ (with $r_s=2$ and $\theta=1$). It is evident that the agreement between the LR-TDDFT results and the PIMC data for $\widetilde F(q,\tau)$ of warm dense hydrogen is very good when $\eta < 0.5~{\rm eV}$. However, the UEG model shows significant deviations from both of these results. As the smearing parameter increases, the LR-TDDFT results computed using the LL method diverge from the PIMC data for hydrogen. This indicates that converged results for the ITCF can be obtained by decreasing the value of $\eta$, despite the noise that arises in the DSF.

We validated the results computed using the LL method by considering the system with $N=14$ protons for which the PIMC data is available. Let us now test the effect of the system size on the DSF. In Fig. \ref{fig:N100}, we compare the DSF results at $r_s=2$ and $T=12.58~{\rm eV}$ computed using 14, 32, and 100 protons at $q = 1.528~\text{Å}^{-1}$ and  $q = 4.584~{\rm \AA^{-1}}$. The overall shape of the DSF, the position of the maximum, and the FWHM are in good agreement for the simulations with different numbers of particles.

\begin{figure}\centering
\includegraphics[width=0.48\textwidth]{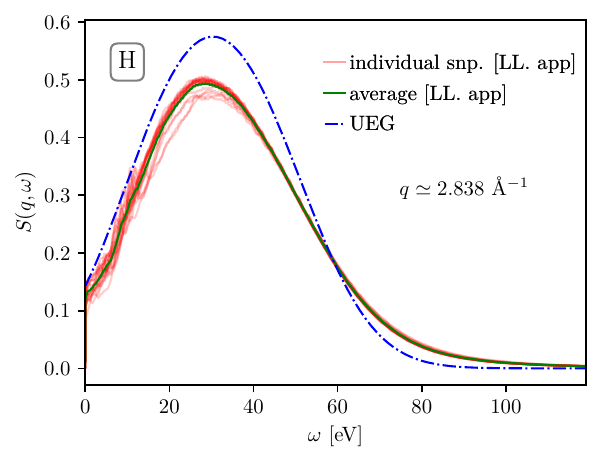}
\includegraphics[width=0.48\textwidth]{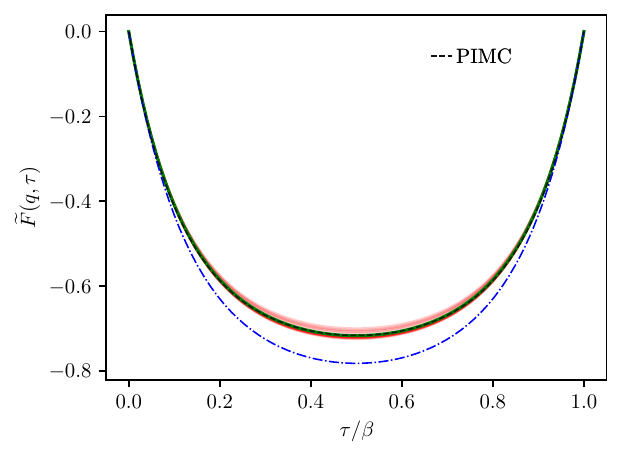}
\caption{\label{fig:UEG_N14_rs2_theta2_switch_probability} 
The simulation results for the DSF (top panel) and shifted ITCF (\ref{eq:tilde_ITCF}) (bottom panel) for warm dense hydrogen at $q\simeq 2.838~{\rm \AA^{-1}}$, $r_s=3.23$ and $T=4.8~{\rm eV}$ with the Lorentzian smearing parameter set to $\eta =0.5~{\rm eV}$. 
The LL method-based LR-TDDFT calculations were performed by averaging over 20 different snapshots of 14 protons. The results are compared with the data computed using the UEG model~\cite{Dornheim_PRL_2020_ESA} and with the PIMC results for warm dense hydrogen \cite{Dornheim_MRE_2024}.
}\label{fig:rs3.23_3qmin}
\end{figure}

\subsection{Heated solid density hydrogen}

Next, we test the LL approach to LR-TDDFT by examining the more challenging case of warm dense hydrogen at solid density, with $ r_s = 3.23 $ and $ T = 4.8 \, \text{eV} $. This regime has stronger coupling between particles and enhanced localization of electrons around protons compared to the case with $r_s=2$ \cite{Dornheim_jcp_2024}.

In Figs. \ref{fig:rs3.23_qmin} and \ref{fig:rs3.23_3qmin}, we present the LL-based LR-TDDFT results for $ S(q, \omega) $ and $ \widetilde F(q, \tau) $ at $ q = 0.94617 \, \text{Å}^{-1} $ (with $ \eta = 0.3 \, \text{eV} $) and $ q = 2.838 \, \text{Å}^{-1} $ (with $ \eta = 0.5 \, \text{eV} $), respectively. The calculations utilize 20 snapshots, and we include the corresponding snapshot-averaged values. The LR-TDDFT results are compared with the PIMC results for warm dense hydrogen from Ref. \cite{Dornheim_MRE_2024}. Additionally, we provide data computed using the UEG model.

From Figs. \ref{fig:rs3.23_qmin} and \ref{fig:rs3.23_3qmin}, it is evident that the LR-TDDFT data averaged over snapshots for $ \widetilde F(q, \tau) $ aligns excellently with the PIMC results at both wavenumbers considered. In contrast, the UEG model shows substantial discrepancies when compared to the LR-TDDFT and PIMC results for hydrogen. We note that $ S(q, \omega) $ and $ \widetilde F(q, \tau) $ in this case exhibit similar dependence on $ \eta $, as discussed in Sec. \ref{sub:rs2}. These findings for solid-density hydrogen further validate the LL approach to LR-TDDFT under warm dense matter conditions, characterized by strong localization of electrons around ions \cite{Dornheim_pre_2023, Moldabekov_jctc_2024}.

\begin{figure}
    \centering
    \includegraphics[width=0.5\textwidth,keepaspectratio]{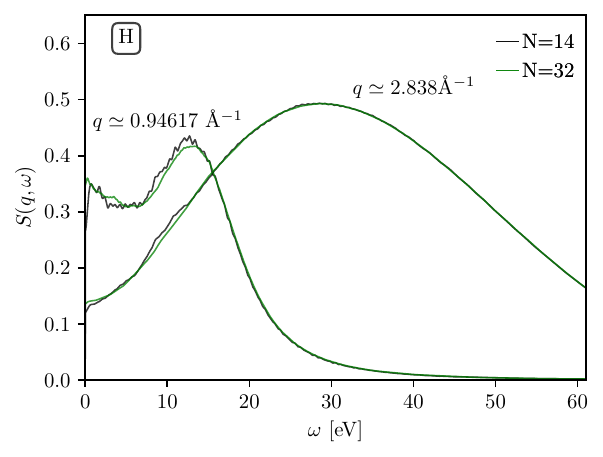}
    \caption{
        Comparison of the DSFs of warm dense hydrogen computed using the LL approach to LR-TDDFT with 14 and 32 particles in the simulation cell at $r_s=3.23$ and $T=4.8~{\rm eV}$. The Lorentzian smearing parameter was set to $\eta=0.3~{\rm eV}$ for $q\simeq 0.946~{\rm \AA^{-1}}$ and to $\eta=0.5~{\rm eV}$ for $q\simeq 2.838~{\rm \AA^{-1}}$.
    }
    \label{fig:rs3.23_N32}
\end{figure}

To investigate the effect of system size on the DSF of hydrogen at $r_s=3.23$, we compare the results for 14 particles with those obtained using 32 particles at $ q = 0.94617 \, \text{Å}^{-1} $ and $ q = 2.838 \, \text{Å}^{-1} $ in Fig. \ref{fig:rs3.23_N32}. The results indicate that increasing the wavenumber reduces the sensitivity of the results to system size as we eventually attain the single-particle limit. The DSF data for 14 and 32 particles show close agreement at  $ q = 2.838 \, \text{Å}^{-1} $. At $ q = 0.94617 \, \text{Å}^{-1} $, the DSF results for 14 and 32 particles exhibit only minor deviations from each other to the left of the DSF maximum, with the data for 14 particles being somewhat noisier.

% \begin{figure}\centering
% \includegraphics[width=0.48\textwidth]{H100_q1.528.pdf}
% \includegraphics[width=0.48\textwidth]{H100_q4.584.pdf}
% \caption{\label{fig:H100} 
%  $N=14$, $r_s=2.0$, $\Theta=1$, $\eta=0.3~{\rm eV}$ and $\eta=3.0~{\rm eV}$
% }
% \end{figure} 

\section{Conclusions}\label{s:end}

We validated and tested the LL method, as implemented in Quantum ESPRESSO (compatible with both CPUs and GPUs), for WDM applications. This step is crucial for expanding the resources available for modeling matter under extreme conditions, given the importance of XRTS measurements as a diagnostic tool in WDM experiments.

To test the LL method, we first considered isochorically heated Al, which features hot electrons in an fcc lattice structure of cold ions. Such transient states are generated and examined in experiments that utilize X-ray-driven heating \cite{Descamps_sciadv, Descamps2020, Matthieu2011, White2014} (also see Ref. \cite{Moldabekov_prr_2024} for detailed discussions). For the fcc Al, we found good overall agreement between the results from the LL method and the solution of Dyson's equation in the standard LR-TDDFT within the PAW formalism. However, we noted differences in the DSF values near the DSF maximum between the two methods. This discrepancy arises from the different approaches in approximating the interaction of core electrons with valence electrons. Specifically, the pseudopotential or PAW setup employed defines the interaction of valence electrons with the crystal lattice, affecting the band structure and, consequently, the DSF as well as EELS \cite{Smith_prb_1986, Schuelke_Si, Schuelke, Si_Weissker, Moldabekov_prr_2024}. Our analysis of the DSF for fcc Al demonstrated that if relatively subtle features of the DSF are of interest, particular care must be taken by cross-validating results using different approaches to TDDFT calculations. This was also pointed out in a recent work by Hentschel \textit{et al.} \cite{Hentschel_POP_2025} considering pseudopotentials with 3 and 11 valence electrons.

Second, we considered warm dense hydrogen as an example of WDM in equilibrium. We tested the results from the LL method by comparing them to LR-TDDFT data computed within the PAW formalism. Additionally, by examining the ITCF derived from the DSF results, we confirmed the high quality of the LL method by benchmarking the data against the PIMC results for hydrogen. This opens the door for further detailed studies of warm dense hydrogen properties such as the DSF, density response functions, and the dynamic dielectric function using the LL method across a wide range of parameters. This research direction is important given the relevance of warm dense hydrogen for ICF and astrophysical applications.

Moreover, our analysis of the effects of pseudopotentials, the number of iterations in the LL method, and the Lorentzian smearing parameter provide a valuable foundation for future applications of the LL method in the WDM regime. 

The primary computational limitation of the standard LR-TDDFT arises from the memory requirements to store matrices on simulation cores \cite{jussi_enkovaara_2012_814432}. This can reach several terabytes when using around $10^3$ bands. When a relatively small number of bands is required, e.g., for wavenumbers $ q \lesssim q_{q_F} $ and small-sized systems (typically containing about $ N \sim 10$ atoms), the standard LR-TDDFT approach can be up to two orders of magnitude more efficient than the LL method. However, in cases involving a large number of bands or atoms, as discussed in this work, employing standard LR-TDDFT becomes computationally unfeasible or highly impractical. In contrast, the LL method has lower memory requirements and does not necessitate empty states, allowing efficient parallelization over thousands of cores. This makes it capable of handling larger wavenumbers, broader frequency ranges, and larger systems.
A detailed discussion of the comparison of the computational costs between the LL method and LR-TDDFT, using bulk diamond as an example, can be found in Ref. \cite{TIMROV2015460}.

We would like to emphasize the utility of the LL method for calculating the DSF at large wavenumbers. In the standard approach to solving Dyson's equation within LR-TDDFT, increasing the wavenumber requires a significant increase in the number of empty bands and memory resources for calculations. Consequently, the standard LR-TDDFT method faces considerable challenges when calculating the DSF at large wavenumbers. In contrast, the LL method does not necessitate empty orbitals, enabling the DSF calculations at large wavenumbers. The XRTS spectra at large q values are particularly important for WDM experiments, as backward scattering geometry measurements are essential for diagnostics at key facilities, including the NIF\cite{Poole_pop_2022, Tilo_Nature_2023} which operates the most powerful laser platform for WDM studies in the world.

In summary, the LL approach to LR-TDDFT is valuable for \textit{ab initio} simulation of WDM properties, with applications in XRTS analysis and the calculation of dielectric properties. For instance, the LL method can be employed to plan and design new experiments. Furthermore, it can be used to test other TDDFT implementations, such as stochastic RT-TDDFT \cite{white_RTTDDFT_2024} and orbital free TDDFT \cite{white_RTTDDFT_2024, PhysRevB.103.245102, PhysRevB.106.115153} for calculating the linear response properties of WDM across various parameters. Finally, we note that LR-TDDFT provides access to a wide range of density response functions (including ideal Kohn-Sham response and random phase approximation with and without local field effects), which is helpful for methodological advancements in fundamental WDM theory.

\begin{acknowledgments}
This work was partially supported by the Center for Advanced Systems Understanding (CASUS), financed by Germany’s Federal Ministry of Education and Research (BMBF) and the Saxon state government out of the State budget approved by the Saxon State Parliament. This work has received funding from the European Research Council (ERC) under the European Union’s Horizon 2022 research and innovation programme
(Grant agreement No. 101076233, "PREXTREME"). 
Views and opinions expressed are however those of the authors only and do not necessarily reflect those of the European Union or the European Research Council Executive Agency. Neither the European Union nor the granting authority can be held responsible for them. This work has received funding from the European Union's Just Transition Fund (JTF) within the project \emph{R\"ontgenlaser-Optimierung der Laserfusion} (ROLF), contract number 5086999001, co-financed by the Saxon state government out of the State budget approved by the Saxon State Parliament. Computations were performed on a Bull Cluster at the Center for Information Services and High-Performance Computing (ZIH) at Technische Universit\"at Dresden and at the Norddeutscher Verbund f\"ur Hoch- und H\"ochstleistungsrechnen (HLRN) under grant mvp00024.%, and on the HoreKa supercomputer funded by the Ministry of Science, Research and the Arts Baden-W\"urttemberg and
%by the Federal Ministry of Education and Research.
\end{acknowledgments}

\section*{Author Declarations}
\subsection*{Conflict of interest}
The authors have no conflicts to disclose.

%%%%%%%%%%%%%%%%%%%%%%%%%%%%%%%%%%%%%%%%%%%%%%%%%%%%%%%%%%%%%%%%%%%%%%%%%%%%%%%%
% literature
%%%%%%%%%%%%%%%%%%%%%%%%%%%%%%%%%%%%%%%%%%%%%%%%%%%%%%%%%%%%%%%%%%%%%%%%%%%%%%%%
\bibliography{bibliography}
\end{document}